\documentclass[preprint,aps,prl,showpacs]{revtex4}

\begin{document}
\title{Phase Transition and Critical Dynamics in Diluted Josephson Junction Arrays}
\author{Young-Je Yun}
\author{In-Cheol Baek}
\author{Mu-Yong Choi}
\affiliation{Department of Physics and Institute of Basic
Science, Sungkyunkwan University, Suwon 440-746, Korea}

\begin{abstract}
Measurements of the $IV$ characteristics of site-diluted
Josephson-junction arrays have revealed intriguing effects of
percolative disorder on the phase transition and the vortex dynamics
in a two-dimensional $XY$ system. Different from other types of
phase transitions, the Kosterlitz-Thouless (KT) transition was
eliminated with the introduction of percolative disorder far below
the percolation threshold. Even after the KT order had been removed,
the system remained superconducting at low temperatures by
establishing a different type of order. Near the percolation
threshold, evidence was found that, as a consequence of the
underlying fractal structure, the critical dynamics of the phase
degrees of freedom persisted down to zero temperature.
\end{abstract}

\pacs{64.60.Cn, 64.60.Ak, 75.10.Hk, 74.81.Fa}

\maketitle

Percolation has been used as an idealized simple model for
understanding a variety of irregular systems. The effects of such a
geometry on phase transitions have been of continuous interest in
condensed matter physics. the Kosterlitz-Thouless (KT) phase
transition \cite{R1} is an essential factor for understanding
various systems in two dimensions, such as superconductors,
easy-plane ferromagnets, crystals, and liquid crystals, which can be
represented by the $XY$ model. In addition, the KT transition is
singular with regard to excitations of topological defects, which
drive the transition from a quasi-long-range ordered phase to a
disordered phase. Nevertheless, the relationship between percolative
disorder and the two-dimensional (2D) phase transition has not been
sufficiently explored and remains unsettled \cite{R2,R3}. Also, of
interest are the effects of the underlying fractal geometry on the
dynamics of the phase degrees of freedom of $XY$ systems near the
percolation threshold. In this paper, we report an experimental
study of the effects of percolative disorder on both the 2D phase
transition and the critical dynamics in proximity-coupled
Josephson-junction arrays (JJA's), which provide a near-ideal
realization of the $XY$ model. Measurements of the current-voltage
($IV$) characteristics demonstrate that the nature of the phase
transition in the 2D $XY$ system is completely altered with
progressive addition of percolative disorder. Near the percolation
threshold, experimental evidence suggests that the underlying
fractal structure results in dynamical criticality being maintained
down to zero temperature.

Experiments were performed on a series of site-diluted square arrays
of  Nb/Cu/Nb Josephson junctions. Nb crosses were periodically
disposed on a Cu film with a lattice constant of 13.7 $\mu $m, a
junction width of 4 $\mu $m, and a junction separation of 1.4 $\mu
$m to form a 400$ \times $600 square lattice. Nb islands were
randomly removed from the lattice to introduce percolative disorder.
Arrays with the concentration of filled sites $p = $ 1, 0.94, 0.86,
0.8, 0.75, 0.7, 0.65, and 0.6 were fabricated. For the samples, the
same random-number seed was used to select the sites to be removed.
The percolation-threshold concentration $p_c$ for the series of
samples, below which no percolating cluster of connected junctions
was formed, was 0.5906 \cite{R4}.

The $IV$ characteristics were measured by employing the
phase-sensitive voltage-signal-detection method with a lock-in
voltmeter and a square-wave current at 23 Hz. The single-junction
critical current $i_c $ and the single-junction coupling energy $J$
($ = \hbar i_c /2e$) were determined by using the de Gennes formula
\cite{R5} for proximity-coupled junctions in the dirty limit to
extrapolate the $i_c $ vs $T$ data at low temperatures and by using
the numerically obtained $I$ vs $dV/dI$ data for site-diluted arrays
in Ref. \cite{R6} to compensate the systematic error arising from
the presence of diluted sites. The $i_c $ vs $T$ data at low
temperatures were obtained from the $I$ vs $dV/dI$ measurements. The
magnetic field or the frustration $f$, defined as the number of flux
quanta per unit cell, was adjusted by using the $R$ vs $f$ curve of
the sample exhibiting distinct resistance minima at integral $f$'s.
The temperature during the measurements was controlled to have
fluctuations $\lesssim 1$ mK. Additional details of the measurements
are given in Ref. \cite{R7}.

Figure\ \ref{fig1} presents some of the results of the $IV$
characteristics measurements on the samples of $p =$ 1, 0.86, 0.7,
and 0.65, with the ambient magnetic fields expelled from the sample
space by the $\mu $-metal shield and solenoid. The measurements were
carried out at 17-19 different temperatures (at intervals shorter
than shown in the figure). The $IV$ curves were obtained by
averaging 15-240 measurements for each current. It is visible that
the $IV$ characteristics change completely with decreasing $p$.
Unlike the $p =$ 1 (or no site disorder) sample experiencing a KT
transition \cite{R8}, the samples with $p =$ 0.7 and 0.65 exhibit
low-temperature $IV$ characteristics with an activated character.
The variation of the voltage signal with temperature both above and
below the transition is significantly slower for $p =$ 0.7 and 0.65
than for $p = 1$ \cite{R9}. For $p=0.86$, the $IV$ characteristics
are more like those of a KT system at $T > T_c$, but they exhibit an
activated character at $T < T_c$. The $IV$ characteristics for $p =$
0.7 and 0.65 are never similar to those of a KT system \cite{R10}.
The low-temperature curves for $p =$ 0.7 and 0.65 can be fitted to
an exponential form $V \sim I \exp[-(I_{T}/I)^{\mu}]$ with $\mu =$
0.9-1.1. The low-temperature curves of an exponential form indicate
that at strong site disorder, the arrays have genuine
superconductivity with long-range phase coherence. The
$d({\log}V)/d({\log}I)$ vs $I/I_{c}$ plots for $p =$ 0.7 and 0.65 in
Fig.\ \ref{fig1}, showing the proposed criterion for a non-KT-type
superconducting transition \cite{R11} to be satisfied, confirm a
finite-temperature non-KT-type superconducting transition for the
strongly disordered samples. The evolution of the $IV$
characteristics with $p$ demonstrates that percolative disorder far
below the percolation threshold alters the nature of the 2D phase
transition in an unfrustrated $XY$ system from a KT-type transition
to a non-KT-type transition.

Figure\ \ref{fig2} shows the $IV$ characteristics of the same
samples with frustration $f = 2/5$. The $IV$ plots for $f = 2/5$
demonstrate that at strong disorder or at lower concentrations of
filled sites, the $IV$ characteristics are affected little by the
imposition of frustration, excluding the $f$-dependent
phase-transition temperature. It is surprising that at strong
disorder, the $IV$ characteristics of the unfrustrated system become
similar to those of frustrated systems, which are known to
experience a melting transition of a vortex solid driven by domain
wall excitations \cite{R12}. The development of the unfrustrated
system into a system similar to frustrated systems with the addition
of percolative disorder is also revealed in the $p$ dependence of
the reduced phase-transition temperature $\tilde{T}_c$
($=T_{c}/(J/k_{B})$), shown in Fig.\ \ref{fig3}. The $T_c$'s in the
figure are the temperatures at which $I \sim V^3$ for systems
exhibiting KT-like $IV$ characteristics or at which the $IV$ curve
becomes straight in a $\log I$-$\log V$ plot for systems exhibiting
non-KT-like $IV$ characteristics \cite{R13}. The $T_c$'s determined
from the $IV$ curves agree, within the error bars, with those
independently determined from the resistance measurements. The
$\tilde{T}_c$ vs $p$ plot shows that the $p$ dependences of
$\tilde{T}_c$ for $f =$ 0 and 2/5, despite the total difference at
weak disorder, become qualitatively similar at strong disorder. The
rapid drop of $T_c$ to zero in the vicinity of $p_c$ is a common
feature of systems with a gap in the excitation spectrum \cite{R14}.
The close resemblance of the $IV$ characteristics and the $p$
dependence of $\tilde{T}_c$ at strong disorder for two systems, one
with and one without frustration, appears to suggest that the
non-KT-type low-temperature order in an unfrustrated $XY$ system at
strong disorder is possibly that of a superconducting vortex solid
with long-range phase coherence, as in frustrated systems.

Although a vortex solid in an unfrustrated array sounds peculiar,
the possibility of its existence is notable because an unfrustrated
array with the KT transition suppressed always contains sufficient
unbound vortices to form a vortex solid. A recent numerical study
\cite{R15} has suggested that in the presence of a random potential
above a critical disorder strength, a breaking of ergodicity due to
a large energy barrier against vortex motion may allow unbound
vortices to form a vortex solid at finite temperatures. In the
equivalent interacting-vortex (or Coulomb-gas) representation of the
$XY$ model, random site dilutions produce a random pinning potential
acting on vortices. Our speculation is that the random pinning
potential eliminates the KT transition far below the percolation
threshold and constrains the unbound vortices to form a
superconducting solid at low temperatures \cite{R16}.

Another important feature of the data presented in this paper is
that the $IV$ characteristics of the unfrustrated system completely
change again near $p_c$ (= 0.5906). As Fig.\ \ref{fig4} reveals, the
low-temperature $IV$ curves shift from exponential-type curves to
power-law-type curves when $p$ changes from 0.65 to 0.6. For $p =$
0.6, an approximate power-law character is maintained even at such
low temperatures as $\tilde{T} = 0.006$, far below $\tilde{T}_c $ (=
0.16) \cite{R17}. The power-law character never weakens, even when
frustration is turned on. In addition, the slope of the isotherm at
$T_c$ in the $\log I$-$\log V$ plot is only ~1.9 for $p=0.6$, less
than it is for $p \gtrsim 0.65$. Since the slope minus one is the
dynamic critical exponent \cite{R13}, the lower slope for $p=0.6$
implies that the critical relaxation in a percolating system is
faster than it is in less-disordered systems. This is contrary to
the general expectation that more diluted sites in a JJA usually
result in slower relaxation of the system. Even if the power-law
$IV$ relation is characteristic of a KT transition, the $f$
independence of the power-law behavior and the dynamic exponent
being much less than 2 indicate that the possibility of a
re-emerging KT transition near $p_c$ is low. The motion of the
vortices created at diluted sites by an injected current does not
explain the observed power-law behavior either. If the vortices
created at the defects were responsible for the power-law behavior
of the $p=0.6$ system, as expected from numerical studies \cite{R18}
of diluted arrays at zero temperature, a similar behavior should be
visible in systems with comparable defect concentrations. The
exponential low-temperature $IV$ curves for $p=$ 0.65, however,
confirm that the contribution of vortices created at defects must be
insignificantly small.

Near the percolation threshold, a self-similar fractal structure
with geometrical inhomogeneities and a divergent percolation
correlation length characterize percolating systems. Therefore, even
at zero temperature where the thermal correlation length, i.e., the
spatial extent of the thermal fluctuations vanishes, the
{\it{spatial}} configuration of the phase angles of the
superconducting islands in the percolating array has a self-similar
critical structure, as at critical temperature. For non-KT-type
superconducting transitions, a power-law $IV$ relation is a salient
feature of {\it{dynamical}} criticality \cite{R13}. The power-law
$IV$ curves far below the transition for $p=$ 0.6 may then indicate
that temperature-independent criticality due to the fractal geometry
applies to the dynamics of the phase angles, as well as to their
spatial configuration. This means that the relaxation time near the
percolation threshold becomes divergent at all temperatures below
$T_c$. It has been suggested \cite{R19} from measurements of the
frequency-dependent complex conductance of a site-diluted 2D JJA
that the fractal structure results in a crossover, at a critical
frequency, of the dynamics of the phase angles from a low-frequency
Euclidean (homogeneous) region dominated by vortices to a
high-frequency fractal (inhomogeneous) region in which the relevant
excitations are localized fractons. The critical frequency was found
from the ac measurements to be $\gtrsim 1$ kHz for arrays comparable
to the present sample. However, the results of the $IV$
characteristics measurements on the present sample appear to
indicate that the low-frequency vortex dynamics is not Euclidean
either, but is in another inhomogeneous region in which the fractal
geometry leads to critical dynamics.

To conclude, the $IV$ characteristics measurements on site-diluted
JJA's have revealed intriguing effects of percolative disorder on
the phase transition and the vortex dynamics in a 2D $XY$ system.
Unlike other types of phase transitions, the KT transition was
eliminated by introduction of percolative disorder far below the
percolation threshold. When the KT order was removed, the system
unexpectedly remained superconducting by establishing a different
type of order with long-range phase coherence until the percolation
threshold was attained. Near the percolation threshold, as a
consequence of the underlying self-similar geometrical structure,
evidence was found for the critical dynamics of the phase degrees of
freedom persisting down to zero temperature.

This work was supported in part by the BK21 program of the
Ministry of Education.

\newpage

\begin{figure}[h!]
\caption{Development of the $IV$ characteristics with the
filled-site concentration $p$ for $f=$ 0. The $I$ vs $V$ isotherms
differ by a temperature interval of 0.1 K. The panels on the right
show the slopes of the isotherms as functions of $I/I_c$. The dashed
lines are drawn to show where the phase transition occurs. For
$p=1$, the dashed line represents the $IV$ curve at the temperature
where $V \sim I^3$. The dashed lines for $p=$ 0.7 and 0.65 are drawn
at the temperatures where the $IV$ curves are estimated to become
straight. For $p=0.86$, $V \sim I^3$ at the transition temperature
where the $IV$ curve becomes straight. See the text for discussions.
} \label{fig1}
\end{figure}

\begin{figure}[h!]
\caption{$I$ vs $V$ and $I/I_c$ vs $d(\log V)/d(\log I)$ plots for
the samples with $p=$0.7 and 0.65 at $f=$ 2/5.}
\label{fig2}
\end{figure}

\begin{figure}[h!]
\caption{$p$ dependence of the reduced phase transition temperature
$\tilde{T}_c$ ($=T_{c}/(J/k_{B})$) for $f=$ 0 and 2/5. The arrow
points to the percolation threshold $p=0.5906$. The inset shows a
portion of a site-diluted array used in the experiment.}
\label{fig3}
\end{figure}

\begin{figure}[h!]
\caption{$IV$ characteristics of the sample with $p=$ 0.6 for $f=$ 0
and 2/5. The $I$ vs $V$ isotherms differ by a temperature interval
of 0.1 K.} \label{fig4}
\end{figure}


\begin{references}
\bibitem {R4}
$p_c$ (=0.5906) for the series of samples was determined for a
specific random-number seed as the filled-site concentration
below which no cluster of connected junctions spanned the current
electrodes of the sample. It is, thus, slightly different from the
theoretical value of 0.5927 for an infinite square array.
\bibitem {R8}
The $IV$ curves of the sample with $p=$1 display the special
features of a KT transition, with the exception of the shoulders at
low currents, which are present in low-temperature curves. The
shoulders in the low-temperature curves are due to the motion of
vortices created by the inhomogeneous residual magnetic field. It is
highly difficult to completely eliminate the inhomogeneous residual
magnetic field when using a $\mu$-metal shield and a solenoid. The
frustration equivalent to the residual magnetic field is less than
0.001, as estimated from the $IV$ data in Ref.\ \cite{R7} for the
same sample in weak magnetic fields. Since the shoulders attributed
to the residual field are observable only in a KT system, where
low-current voltage signals below the transition are very small, the
presence of such shoulders may be regarded as another indication of
a KT transition.
\bibitem {R9}
The horizontal distances between adjacent isotherms are much
narrower for $p=$0.7 and 0.65 than for $p=$1.
\end{references}
\end{document}